\documentclass[12pt]{article}
\usepackage{geometry}
\usepackage{textcomp,booktabs}
\usepackage[usenames,dvipsnames]{color}
\usepackage{colortbl}
\usepackage{epsfig,graphicx}
\usepackage{subfig}
\usepackage{float}
\usepackage{setspace,multirow}
\usepackage{amsmath,amsthm,amsfonts}
\usepackage{amssymb,enumerate}
\usepackage{mathrsfs}
\usepackage{setspace,paralist,color}
\usepackage{bm,verbatim}
\usepackage{lineno}
\usepackage{multicol}
\usepackage{rotating}
\usepackage{MnSymbol}
\usepackage{threeparttable}
\usepackage[titletoc]{appendix}
\usepackage{titlesec}
\usepackage{makecell}
\usepackage[colorlinks,citecolor=blue,linkcolor=black]{hyperref}
\usepackage{algorithm,algorithmic}

\definecolor{mygray}{gray}{.9}
\definecolor{mypink}{rgb}{.99,.91,.95}
\definecolor{mycyan}{cmyk}{.3,0,0,0}
\allowdisplaybreaks

\def\E{\hbox{E}}

\def\T{{ \mathrm{\scriptscriptstyle T} }}

\def\E{\hbox{E}}

\usepackage{xr}

\begin{document}

\begin{singlespace}
		
		\title{A Review and Classification of Model Uncertainty
		}
		
		\author{Guangyuan Cui, Yuting Wei, Xinyu Zhang
		}
		
		\author{Guangyuan Cui$^{1,3}$, Yuting Wei$^{2}$, and Xinyu Zhang$^{3}$
			\\{ \footnotesize $^{1}$ Department of Decision Analytics and Operations, City University of Hong Kong}\\
			{ \footnotesize $^{2}$ School of Mathematics and Statistics, Nanjing University of Information Science and Technology}\\
			{\footnotesize $^{3}$ Academy of Mathematics and Systems Science, Chinese Academy of Sciences}\\
		}
		
		\date{}

\maketitle

\begin{abstract}
	\noindent 
	Model uncertainty is a crucial issue in statistics, econometrics and machine learning, yet its definition remains ambiguous and is subject to various interpretations in the literature. So far, there has not been a universally accepted definition of model uncertainty. We review different understandings of model uncertainty and categorize them into three distinct types: uncertainty about the true model, model selection uncertainty, and model selection instability. We further offer interpretations and examples for a better illustration of these definitions. We also discuss the potential consequences of neglecting model uncertainty in the process of conducting statistical inference, and provide effective solutions to these problems. Our aim is to help researchers better understand the concept of model uncertainty and obtain valid statistical inference results on the premise of its existence.\newline
\end{abstract}
\end{singlespace}

\section{Introduction}\label{Introduction}
Model uncertainty is crucial across many fields, such as statistics, economics and machine learning, and it plays a pivotal role in statistical modeling, particularly in the study of statistical inference and asymptotic distribution. However, most existing studies often ignore the presence of model uncertainty, not to mention exploring its effect on the present statistical results. On the other hand, the definitions of model uncertainty remain ambiguous and controversial among the relevant literature. Therefore, it is essential to provide a detailed classification of the definitions and subsequent analysis for model uncertainty. 

To the best of our knowledge, existing definitions of model uncertainty are generally categorized into three types: \textit{uncertainty about the true model} (i.e., the data-generating process), \textit{model selection uncertainty}, and \textit{model selection instability}. Uncertainty about the true model encompasses uncertainty about the functional form, distribution assumptions, and variable involved in the true model. This type of definition is widely mentioned in Madigan and Raftery \cite{Madigan1994}, Kass and Raftery \cite{Kass1995}, Chatfield \cite{Chartfield1995}, Draper \cite{draper1995}, Laskey \cite{Laskey1996}, Hoeting et al. \cite{Hoeting1999}, Cairns \cite{CAIRNS2000}, Clyde and George \cite{clyde2004}, Brock et al. \cite{Brock2007}, Moghadam \cite{Amini2012} and Hansen \cite{LHansen2014}, among others. The second type of definition refers to the uncertainty embedded in the observed model selection results. Due to the randomness of sampling, a single model can hardly describe the data generating process underlying the observed data, and different models can be selected for the same problem, even when applying the same selection procedure to different batches of data. This uncertainty is a reflection of the fact that the true model is unknown. This opinion is shared by Norris and Pollock \cite{Norris1996}, Hjort and Claeskens \cite{Hjort2003}, Ishwaran and Rao \cite{Ishwaran2003} and Allenbrand and Sherwood \cite{Allenbrand2023}. The third definition of model uncertainty is model selection instability, and this instability exhibits itself when a slight change in the original data leads to a different selected model despite using the same selection method (Hong and Wang, \cite{Hongandwang2021}, Yuan and Yang, \cite{Yuan2005}, Chen et al., \cite{Chen2007}).

Drawing upon the existing literature on model uncertainty, this article presents a review on model uncertainty, explaining various interpretations of model uncertainty derived from multiple perspectives. Furthermore, it outlines several approaches to handling the consequences of ignoring different types of model uncertainty. 

The remainder of this paper is organized as follows. Section \ref{Uncertainty towards true model} introduces the uncertainty about the true model and several related concepts. Methods for addressing such uncertainty, including Bayesian model averaging (BMA) and Frequentist model averaging (FMA), are also mentioned. Section \ref{section 3} discusses the model selection uncertainty and offers a summary of the consequences for the behaviour of conducting statistical inference based on a selected candidate model. We also give an introduction to the post-selection inference, which providing valid statistical inference in face of the model selection uncertainty and these negative results. Section \ref{section 3} also introduces the concepts of model confidence sets as supplementary tools to measure the amount of model selection uncertainty. Section \ref{section 4} provides an explanation of model selection instability and several techniques for its measurement. Section \ref{section 6} offers a summary of this paper.

\section{Uncertainty about the true model}\label{Uncertainty towards true model}
A prevailing opinion among the literature tends to define model uncertainty as the uncertainty about the true model (the underlying data-generating process). This uncertainty often manifests in various forms, including the model structure, variables under consideration and distributional assumptions associated with the true model. We provide a summary of the relevant literature as follows.

\subsection{Literature review}
The literature that conceptualizes model uncertainty as uncertainty about the true model can be generally categorized into two types depending on the different ways of illustration. The first type explicitly defines model uncertainty as the uncertainty about the true model (Kass and Raftery, \cite{Kass1995}; Laskey, \cite{Laskey1996}; Brock et al., \cite{Brock2007}), while the second shares the same perspective through inexplicit explanations and examples (Madigan and Raftery, \cite{Madigan1994}; Draper, \cite{draper1995}; Chatfield, \cite{Chartfield1995}; Hoeting et al., \cite{Hoeting1999}; Cairns, \cite{CAIRNS2000}; Clyde and George, \cite{clyde2004}; Moghadam, \cite{Amini2012}).

Specifically, according to Kass and Raftery \cite{Kass1995} and Laskey \cite{Laskey1996}, model uncertainty usually refers to the uncertainty about the appropriate functional form of the model, the correct set of variables or predictors that should be included, and the validity of underlying assumptions, such as the choice of probability distributions or conditional independence relationships. By Laskey \cite{Laskey1996}, this uncertainty is particularly salient in situations where there is no single, universally accepted model for the problem under investigation. Brock et al. \cite{Brock2007} define model uncertainty more broadly as the uncertainty embedded within the data-generating process, and illustrate this point with the example of uncertainty in determining the relevant information set for evaluating policy effects. To sum up, these perspectives highlight the challenges and subjective judgments involved in translating real-world complexity into a formal statistical model, even before considering model fitting or selection based on the observed data.

Another line of research describes model uncertainty with no definitions, but by providing explanations or examples, through which model uncertainty is conveyed as the uncertainty about the true model. This perspective acknowledges that uncertainty extends the scope of parameter values within a selected model structure to encompass the model structure itself (Draper, \cite{draper1995}). The structural uncertainty may concern the relationship between specific variables (Madigan and Raftery, \cite{Madigan1994}) or the overall model form, and this type of uncertainty persists even for a prespecified model, as this model can potentially be misspecified or merely an approximation to the true model (Chatfield, \cite{Chartfield1995}). However, despite its great importance, structural uncertainty often receives less attention than the parameter uncertainty (Draper, \cite{draper1995}).

The degree of uncertainty regarding the true model can be quantified before and after observing data by separately examining the prior distributions of each model and the parameters therein and the posterior probabilities distributed across the candidate models conditional on the observed data, as suggested by Clyde and George \cite{clyde2004}. This perspective is somewhat confirmed by Hoeting et al. \cite{Hoeting1999}, which states that a substantial amount of uncertainty is believed to exist when there is no single model attaining a high posterior probability, and the probability mass is spread thinly across many contenders. Such a case suggests the data and prior information have yielded limited insight into the true underlying structure. On the contrary, the presence of a dominant model often points towards greater certainty. In addition, Hoeting et al. \cite{Hoeting1999} note that uncertainty is likely to be greater when the set of potential models is iteratively developed during the study instead of being fixed beforehand. This reflects less prior knowledge on the studied problem. Note that one way to account for model uncertainty is to consider averaging over a smaller subset of candidate models with high posterior probabilities (Madigan and Raftery, \cite{Madigan1994}).

Researchers have also identified various sources of uncertainty about the true model or other related forms of uncertainty. For instance, Cairns \cite{CAIRNS2000} outlines three principal sources for the uncertainty of the quantity of interest, which can be summarized as inherent stochasticity within a model, the parameter uncertainty and the uncertainty in the model behind the observed data. Furthermore, depending on the level of knowledge of the underlying data generating process, the uncertainty about the true model manifests in three different forms. The high level of understanding of the true model corresponds to the case where the true model is guaranteed to be an unknown member of a general class of models, and the lowest level arises when candidate models serve merely as a proxy for a complex reality for which researchers have limited or no prior information. According to Moghadam \cite{Amini2012}, model uncertainty consists of three types of forms: theory uncertainty, heterogeneity uncertainty, and functional form uncertainty. Theory uncertainty primarily focuses on which variables should be included as determinants in the model. Heterogeneity uncertainty considers whether the relationship between variables remains the same across different individuals. Lastly, functional form uncertainty is related to the form in which variables exist in the true model, whether linear or nonlinear.

Finally, it is worth mentioning that the concept of model uncertainty in this section bears certain resemblance to some other notions, such as \textit{epistemic uncertainty} and \textit{model ambiguity}. As summarized in Hullermeier and Waegeman \cite{Hullermeier2021}, epistemic uncertainty refers to the uncertainty due to the ignorance or lack of knowledge about the true model, and it is crucially important in machine learning. This uncertainty is in principle reducible with the help of additional information. Specifically, there are two sources of uncertainty that eventually constitute the epistemic uncertainty. The first type, named as model uncertainty by Hullermeier and Waegeman \cite{Hullermeier2021}, represents the discrepancy between the true model $f^*$ and the best model $g^*$ among a class of models to be fit $\mathcal{M}$, while the second type measures the uncertainty about how well the model $\hat{g}$ produced by a learning algorithm approximates $g^*$. It is obvious that the first source of uncertainty corresponds to the uncertainty about the true model, as both intend to describe the uncertainty surrounding the target true model. Ideally, this uncertainty would decrease as $\mathcal{M}$ continues to expand and contain a larger class of models.
On the other hand, the concept of model ambiguity was originally proposed in a series of works by Hansen and Sargent \cite{LHansen2012}, Hansen \cite{LHansen2014}, Hansen and Marinacci \cite{LHansen2016}, and Hansen and Sargent \cite{LHansen2022}, among others, and it is defined as the uncertainty about which model to choose among a candidate model set, each of which can potentially be the true model. Then, an event is chosen with a certain probability under the selected model to produce the final observed outcome. Together, these two steps form the generating process of the observed data. To address this ambiguity, Hansen \cite{LHansen2014} suggests assigning subjective probabilities to each candidate model, which provides a way to average the outcomes from all candidate models for subsequent analysis.

\subsection{Methods to handle the uncertainty about the true model}\label{Section 2.2}
Due to the uncertainty about the true model, it often suffices to establish the working model by applying parametric or nonparametric methods to observed data before conducting inference and making predictions. As is widely acknowledged that there exists no model that is universally superior to its alternatives for any dataset and goal, it is important to consider a group of candidate models and then select the most appropriate one based on a prespecified criterion by model selection, and subsequent inference and analysis are conducted based on this selected model. This process is widely adopted in many statistical problems. 

However, the selected model is often misused as the true model for the inference part, which typically ignores the uncertainty about the true model (Chatfield, \cite{Chartfield1995}) as well as the inherent uncertainty in the model selection process (Buckland et al., \cite{buckland1997}; Hoeting et al., \cite{Hoeting1999}; Lubke et al., \cite{Lubke2017assessing}). This could eventually lead to serious consequences. For example, there exists a distortion between the estimator of the parameter of interest that takes account of the randomness and uncertainty in model selection and the one that does not (Zhang et al., \cite{Zhang2022psireview}). Secondly, based on the distorted sampling distribution, the conditional (on the selected model) and unconditional coverage probabilities of the resulting confidence intervals are lower than their nominal levels. A summary of these consequences is further demonstrated in Section \ref{Section 2}. 

A number of methods have been proposed to deal with the invalid statistical inference results caused by neglecting the uncertainty about the true model. Chatfield \cite{Chartfield1995} suggests using the nonparametric method to tackle uncertainty about the true model as fewer assumptions about the model are required compared to parametric methods. In the parametric framework, BMA is widely acknowledged by Draper \cite{draper1995}, Hoeting et al. \cite{Hoeting1999} and Clyde and George \cite{clyde2004} as a useful tool to handle this type of uncertainty. Furthermore, some studies have explored the use of FMA to tackle uncertainty about the true model (Hjort and Claeskens, \cite{Hjort2003}; Zhang and Liu, \cite{Zhang2019mmajma}; Mitra et al., \cite{Mitra2019}; Zhang et al., \cite{zhang2020}; Yu et al., \cite{Yudalei2024}). Both methods are dedicated to performing valid statistical inference in face of the uncertainty about the true model and the possible negative consequences.

\subsubsection{Bayesian model averaging}
According to Draper \cite{draper1995}, Bayesian model averaging can effectively address the structure uncertainty about the true model. Under this framework, the entire model $M=(S,\theta)$ is considered random, where $M$ encompasses both the model structure $S$ and the unknown parameter $\theta$ within $S$. To account for the uncertainty about the true model, it is natural to integrate over the candidate model set $\mathcal{M}$, as shown by
\begin{equation} \label{Posterior probability}
	p\left(y \mid x, \mathcal{M}\right)=\int_{\mathcal{M}} p(y \mid x, M) p(M \mid x) \mathrm{d} M
	=\iint p(y \mid x, \theta, S) p(\theta, S \mid x) \mathrm{d} \theta \mathrm{d} S.
\end{equation}
Here, $y$ and $x$ represent the response and covariates, respectively. From equation \eqref{Posterior probability}, it can be seen that BMA is a novel extension beyond the standard statistical approach that deals with parameter uncertainty, as BMA also considers the uncertainty in the model structure. Meanwhile, the calculation of $p(y \mid x, \mathcal{M})$ involves weighting $p(y \mid x, \theta, S)$ by $p(\theta, S\mid x)$ and then integrating over $\theta$ and $S$, and $p(\theta, S\mid x)$ serves as a measurement of the uncertainty level about $(S, \theta)$ being the true model conditional on $x$. Notably, model selection is actually a special case of BMA, where the selected model is treated as predetermined. In this case, $\mathcal{M}$ only contains a non-random result of model selection, and equation \eqref{Posterior probability} is reformulated as
\begin{equation} \label{Posterior 1}
	p\left(y \mid x, \mathcal{M}\right)=\int p(y \mid x, \theta^*, S^*) p(\theta^*, S^* \mid x) \mathrm{d} \theta^*,
\end{equation}
where $S^*$ is the selected model, and $\theta^*$ represents the parameter within $S^*$. Once again, \eqref{Posterior 1} shows that conditioning on a single selected model fails to address the uncertainty about the model structure of the true model.

By taking extra considerations into the uncertainty about $S$, the traditional Bayesian formulation is modified as 
\begin{eqnarray}\label{Con 1}
	\left\{\begin{array}{ccc}
		\theta & \sim & p(\theta) \\
		(x \mid \theta) & \sim & p(x \mid \theta) \\
		(y \mid x, \theta) & \sim & p(y \mid x, \theta)
	\end{array}\right\} \longrightarrow
	\left\{\begin{array}{ccc}
		S & \sim & p(S) \\
		(\theta \mid S) & \sim & p(\theta \mid S) \\
		(x \mid \theta, S) & \sim & p(x \mid \theta, S) \\
		(y \mid x, \theta, S) & \sim &p(y \mid x, \theta, S)
	\end{array}\right\}.
\end{eqnarray}
Typically, it requires an explicit form of $p(\theta, S \mid x)$ to obtain equation \eqref{Posterior probability}, which in turn demands the specification of $p(S)$ defined in \eqref{Con 1}. As mentioned by Draper \cite{draper1995}, a solution is to utilize the model expansion technique for determining $p(S)$. Specifically, this method expands the structure of the candidate models from a single case to multiple cases, depending on the features of the study.
For the discrete case, assuming that $\mathcal{M}=\{S_1, \ldots, S_M\}$ is the expanded model set containing $M$ candidate models, equation \eqref{Posterior probability} is reformulated as
\begin{eqnarray*} 
	p(y \mid x,\mathcal{M})&=&\sum_{m=1}^{M} \int p\left(y \mid x, S_m, \theta_m\right) p\left(S_m, \theta_m \mid x\right) \mathrm{d} \theta_m\nonumber\\
	&=&\sum_{m=1}^{M} p\left(S_m \mid x\right) p\left(y \mid x, S_m\right).
\end{eqnarray*}

Another interesting point observed by Draper \cite{draper1995} is the following decomposition for $\E(y \mid x,\mathcal{M})$ and $V(y \mid x,\mathcal{M})$:
\begin{eqnarray}
	\E(y \mid x,\mathcal{M})&=&\E_{S}\{\E(y \mid x, S)\}=\sum_{m=1}^{M} \pi_m \mu_m, \label{mean decomposition}\nonumber\\
	V(y \mid x,\mathcal{M})&=&\E_{S}\{V(y \mid x, S)\}+V_{S}\{\E(y \mid x, S)\} \nonumber\\
	&=&\sum_{m=1}^{M} \pi_m \sigma_m^{2}+\sum_{m=1}^{M} \pi_m\left(\mu_m-\mu\right)^{2},\label{variance decomposition}
\end{eqnarray}
where $\mu_m$ and $\sigma_m^2$ represent the mean and variance of $y$ conditional on $x$ and $S_m$, and $\pi_m = p(S_m \mid x)$ denotes the posterior probability of $S_m$. Equation \eqref{variance decomposition} is regarded as the foundation for measuring the uncertainty about the true model, in which the overall uncertainty about $y$ is decomposed into $\E_{S}\{V(y \mid x, S)\}$ and $V_{S}\{\E(y \mid x, S)\}$, where $\E_{S}\{V(y \mid x, S)\}$ represents the average uncertainty given respective choices of model structure and $V_{S}\{\E(y \mid x, S)\}$ stands for the uncertainty in $y$ emerging from the structural uncertainty.

One crucial step of implementing BMA is to choose prior distributions for both the candidate models and the parameters therein, as it characterizes the degree of uncertainty surrounding the true model structure as well as the population parameter value. Priors for parameters may contain conjugate priors, non-informative priors, weakly informative priors, hierachical priors, shrinkage priors and mixture priors, among others (see Gelman et al., \cite{gelman1995bayesian} for a detailed description). On the other hand, common priors for models generally contain uniform, binomial and Beta-binomial model priors. The uniform prior assign equal prior probabilities to all models in the candidate model set $\mathcal{M}$. The binomial prior assigns $p(S_m)=\theta^{p_m}(1-\theta)^{p-p_m}$ for the model $S_m$, where $\theta\in(0,1)$ is a constant, $p_m$ is the dimension of the model $S_m$ and $p$ is the total number of covariates. A slight modification to the binomial prior induces the Beta-binomial prior (Ley and Steel, \cite{ley2009effect}), which further assumes $\theta$ to come from a Beta distribution, i.e., $\theta\sim Beta(a,b)$ with $a, b>0$. 

Despite the fact that BMA is a good approach to accounting for the uncertainty about the model structure, thereby taking full account of the uncertainty about the true model, the use of BMA still suffers from certain drawbacks: 
\begin{enumerate}
	\item The true model is required to be in the candidate model set, but this assumption is challenging to verify in practice because the true model often remains unknown. In cases where all candidate models are misspecified, the generalization performance of BMA is only suboptimal (Masegosa, \cite{masegosa2020learning}).
	\item Conflicts may arise between the settings of prior distributions for the parameters of interest under each candidate model (Hjort and Claeskens, \cite{Hjort2003}).
	\item The selection of prior distributions for candidate models and the parameters therein heavily relies on subjective experience (Draper, \cite{draper1995}; Hoeting et al., \cite{Hoeting1999}), and different prior distributions often yield significantly different BMA results.
	\item It typically involves calculating Bayes factors in order to compute the posterior probabilities of the models and parameters therein for the BMA method, and this may pose a heavy computational burden as this process requires complex integration over the model and parameter space. Solutions to this problem include using the \textit{Occam's window} criterion (Madigan and Raftery, \cite{Madigan1994}) to find a subgroup of models with high posterior probabilities or the Markov Chain Monte Carlo method to find a simple approximation to the integration.
\end{enumerate}

\subsubsection{Frequentist model averaging}
Frequentist model averaging seeks to incorporate information from each candidate model using a set of frequentist methods, thereby taking account of the uncertainty about the true model. The central focus of FMA is to determine a weight choice criterion that aligns with the target problem, and it can be roughly categorized into the following types depending on the way of parameter estimation and determining model weights: boosting (Freund, \cite{Freund1995boosting}), bagging (Breiman, \cite{Breiman1996bagging}), information criteria weighting (smoothed AIC and smoothed BIC, Buckland et al., \cite{buckland1997}), adaptive regression by mixing (Yang, \cite{Yang2001}; Yuan and Yang, \cite{Yuan2005}; Ghosh and Yuan, \cite{Gosh2009}) and optimal model averaging (Hansen, \cite{Hansen2007}; Hansen and Racine, \cite{Hansen2012}; Zhang et al., \cite{zhang2016}; Zhang et al., \cite{zhang2020}). See Peng et al. \cite{Peng2024} for a detailed summary of frequentist model averaging methods. Note that optimal model averaging seeks to pursue a convex combination of estimation or prediction results from all candidate models, and its performance is guaranteed by asymptotic optimality (Li, \cite{li1987}; Andrews, \cite{Andrews1991}; Hansen, \cite{Hansen2007}) when all models are misspecified.

A number of works are dedicated to developing statistical inference for the averaging results. For instance, by assuming perfect correlation between estimators from different models, Buckland et al. \cite{buckland1997} derive the form of the variance for $\hat \mu=\sum_mw_m\hat \mu_m$, i.e.,
\begin{equation}\label{Con 2}
	\operatorname{var}(\hat{\mu})=\left\{\sum_m w_m \sqrt{\operatorname{var}(\hat{\mu}_m \mid \mu_m)+(\hat \mu_m-\hat \mu)^2}\right\}^2,
\end{equation}
where $w_m$ and $\hat \mu_m$ represent the weight and the estimator of $\mu$ under the $m$th model, respectively. This variance form \eqref{Con 2} is later used by Burnham and Anderson \cite{Burnham2002} to construct a confidence interval for $\mu$:
\begin{equation}\label{Burnham Confidence Interval}
	\mu \in \hat\mu \pm z_{1-\alpha/2}\sqrt{\hat{\mbox{var}}(\hat \mu)},
\end{equation}
where \(z_{1-\alpha/2}\) is the \((1-\alpha/2)\)th percentile of a standard Gaussian distribution, and \(\hat{\operatorname{var}}(\hat{\mu})\) is an estimate of \eqref{Con 2}. There are two choices of \(w_m\), including the smoothed AIC and the frequency of selecting the \(m\)th model among a series of repeated experiments involving resampling and a model selection technique.
However, the variance form \eqref{Con 2} is proved incorrect by Hjort and Claeskens \cite{Hjort2003}, as it is founded on the unreasonable assumption of the correlation between different estimators. Moreover, the coverage probabilities of the induced confidence intervals, including \eqref{Burnham Confidence Interval} and that in Allenbrand and Sherwood \cite{Allenbrand2023}, are shown to be biased and lower than their nominal confidence level according to the theoretical results in Hjort and Claeskens \cite{Hjort2003}.

Hjort and Claeskens \cite{Hjort2003} establish the asymptotic distribution of the averaging estimator and corresponding confidence intervals under the local misspecification framework. Under this framework, a generic postulated density for the observed data $Y_1,\ldots,Y_n$ is $f(y,\theta,\gamma)$, where $\theta\in \mathbb{R}^p$ corresponds to the compulsory part of the true model, and $\gamma\in \mathbb{R}^q$ is considered optional for building different models. The data generating process is 
\begin{equation}\label{Con 10}
	f_{\operatorname{true}}(y)=f(y,\theta_0,\gamma_0+\delta/\sqrt{n}),
\end{equation}
where $\theta_0$ and $\gamma_0$ are the true values of $\theta$ and $\gamma$, respectively, and $\delta = (\delta_1,\ldots,\delta_q)$ represents the extent of model departure in all directions. Each model normally takes the form of $f(y,\theta,\gamma_S)$, with $S\subset\{1,\ldots,q\}$. For each element of $\gamma_S$, $\gamma_{S,j}=\gamma_j$ if $j\in S$, and $\gamma_{S,j}=\gamma_{0,j}$ if $j\notin S$. On this basis, for the parameter of interest $\mu$, its true value $\mu_{\mathrm{true}}$ can be written as a function of $\theta_0$, $\gamma_0$ and $\delta$, i.e., $\mu(\theta_0,\gamma_0+\delta/\sqrt{n})$, and an averaging estimator is given by $\hat \mu = \sum_mc(m\mid D_n)\hat \mu_m$, where $c(m\mid D_n)$ denotes the model weight assigned to the $m$th model conditional on $D_n=\sqrt{n}(\hat\gamma_{\mathrm{full}}-\gamma_0)$, and $\hat\gamma_{\mathrm{full}}$ is the maximum likelihood estimate of $\gamma$ under the model with $S=\{1,\ldots,q\}$ (the full model). Hjort and Claeskens \cite{Hjort2003} manage to derive a non-normal asymptotic distribution of $\sqrt{n}(\hat \mu-\mu_{\mathrm{true}})$, based on which a confidence interval for $\mu$ is constructed as follows:
\begin{equation*}
	\mu\in \left[\hat\mu-\hat{\omega}^{\T}\{D_n-\hat\delta(D_n)\}/\sqrt{n}\right] \pm u\hat\kappa/\sqrt{n}.
\end{equation*}
Here, $u$ is a normal quantile. Estimators $\hat{\omega}$ and $\hat\kappa$ are consistent to their true values $\omega$ and $\kappa$, and $\hat\delta(\cdot)$ is a function, which also represents an averaging estimator for $\delta$. For detailed forms of these quantities, please refer to Hjort and Claeskens \cite{Hjort2003}. This confidence interval is shown to asymptotically achieve its nominal coverage probability under certain conditions. 
The local misspecification framework is further adapted to more complicated problems, such as the Cox hazard regression model (Hjort and Claeskens, \cite{Hjort2006Cox}), generalized additive partial linear models (Zhang and Liang, \cite{zhang2011liang}) and least squares averaging estimators (Liu, \cite{Liu2015}). 
However, one drawback of this framework lies in its assumption that all candidate models are within an $O({n}^{-1/2})$ neighborhood of the true model, as implied by \eqref{Con 10}. Mitra et al. \cite{Mitra2019} make an improvement to this framework by altering the true model from \eqref{Con 10} to $f_{\operatorname{true}}(y)=f(y,\theta_0,\gamma_0)$, thereby relaxing the assumption that restricts the distance between candidate models and the true model. The asymptotic normality property of the corresponding averaging estimator is established accordingly, but the model weights are held non-random in this case.

In addition to the local misspecification framework, another field of study (Zhang and Liu, \cite{Zhang2019mmajma}; Yu et al., \cite{Yudalei2024}) focuses on deriving the asymptotic distribution for optimal model averaging estimators and simulation-based confidence intervals are incidentally developed for the coefficients. Focusing on the coefficients $\theta$ in the linear regression setting, Zhang and Liu \cite{Zhang2019mmajma} obtain the asymptotic distributions of the Mallows model averaging (MMA, Hansen, \cite{Hansen2007}) and Jackknife model averaging (JMA, Hansen and Racine, \cite{Hansen2012}) estimators, denoted as $\hat \theta(\hat w_{\text{MMA}})$ and $\hat \theta(\hat w_{\text{JMA}})$, respectively. For instance, under certain conditions, $\hat \theta(\hat w_{\text{MMA}})$ asymptotically converges to a nonstandard distribution 
\begin{equation}\label{Con 11}
	\sqrt{n}\left\{\hat{{\theta}}\left(\hat{{w}}_{\mathrm{MMA}}\right)-{\theta}\right\} \stackrel{d}{\longrightarrow}\sum_{m=1}^{M} \tilde{\lambda}_{\mathrm{MMA}, m} {V}_m {Z},
\end{equation}
where $M$ is the number of just-fitted and over-fitted models, $V_m$ is a nonrandom matrix and $Z$ follows a Gaussian distribution. Note that $\{\tilde \lambda_{\text{MMA},k}\}_{m=1}^{M}$ is the solution to a quadratic programming problem over $\{{\lambda} \in[0,1]^{M}: \sum_{m=1}^{M} \lambda_m=1\}$, which has no closed form, so it is infeasible to construct confidence intervals for $\theta$ simply based on the limiting distribution in \eqref{Con 11}. To this end, a simulation-based confidence interval is developed for each element of $\theta$:
\begin{equation}\label{Con 33}
	\left[\hat{\theta}_j(\hat{{w}}_{\mathrm{MMA}})-n^{-1 / 2} \hat{q}_j(1-\alpha / 2), \hat{\theta}_j(\hat{{w}}_{\mathrm{MMA}})-n^{-1 / 2} \hat{q}_j(\alpha / 2)\right].
\end{equation}
The key step in computing \eqref{Con 33} is determining $\hat q_j(\alpha/2)$ and $\hat q_j(1-\alpha/2)$, which represent the $(\alpha/2)$th and $(1-\alpha/2)$th quantiles of a set of bootstrap estimates for the $j$th element of $\sum_{m=1}^{M} \tilde{\lambda}_{\mathrm{MMA}, k} {V}_m{Z}$ obtained from $B$ bootstrap samples. Moreover, numerical studies of Zhang and Liu \cite{Zhang2019mmajma} suggest that the coverage probability of \eqref{Con 33} would approach its nominal level as $n$ and $B$ go to infinity. An extension to the generalized linear models is made by Yu et al. \cite{Yudalei2024}, and similar theoretical results were obtained accordingly. In addition, the asymptotic distribution of the Parsimonious averaging estimators for the coefficients of regression is established by Zhang et al. \cite{zhang2020}, that is, ${\alpha}^{{\T}}({\text{X}^{*}} ^{\T} \text{X}^*)^{1 / 2}\{\hat{{\theta}}(\hat{{w}})_{\mathcal{A}}-{\theta}_{\mathcal{A}}\}\stackrel{d}{\longrightarrow} \operatorname{N}\left(0, \sigma^2\right)$. Here, $\alpha$ is a vector of norm $1$, the index set $\mathcal{A}$ corresponds to non-zero coefficients in the most parsimonious model, and $\mathrm{X}^*$ is the corresponding design matrix.

\section{Model selection uncertainty}\label{section 3}

Despite the importance of model selection, this process is often fraught with uncertainty. Model selection uncertainty arises from the inherent difficulty in definitively identifying the single best model based on a prespecified criterion and the observed data. Due to the natural sampling variability, a different model might be selected as the preferred one if a new sample were drawn from the same population. The main cause of this uncertainty is the discrepancy between the sample and population distributions, and this uncertainty is expected to decrease as the discrepancy decreases with an increasing sample size (Lubke et al., \cite{Lubke2017assessing}). In addition to the randomness of the sample, it also extends to the plausibility of the considered models and the parameter estimates therein. In essence, model selection uncertainty exists whenever there exist reasonable alternative models for consideration that lead to different characterizations of the underlying data generating process. 

To the best of our knowledge, a number of studies provide an implication that model uncertainty primarily refers to the model selection uncertainty. For instance, Ishwaran and Rao \cite{Ishwaran2003} interpret model uncertainty as the additional variance generated when selecting among a set of unknown parameters. This additional variance is essentially related to the specific model selection method, the candidate model set and the observed data, thereby implying our statement. Moreover, the above finding is also confirmed by Norris and Pollock \cite{Norris1996}, which focuses on the variance of the estimator $\hat\theta_{\hat m}$ that takes account of the uncertainty in the selection outcome $\hat m$. They argue that $\mbox{var}(\hat{\theta}_{\hat m})$ would capture the uncertainty caused by the model selection process, whereas $\mbox{var}(\hat{\theta}_m)$ does not account for this uncertainty, which naturally leads to the result $\mbox{var}(\hat{\theta}_{\hat{m}})\neq \mbox{var}(\hat{\theta}_m)$. Moreover, the uncertainty arising from the model selection process helps constitute the uncertainty in the parameter estimates or quantities returned by any candidate model (Allenbrand and Sherwood, \cite{Allenbrand2023}).

\subsection{Consequences of ignoring model selection uncertainty}\label{Section 2}
As mentioned in Section \ref{Section 2.2}, conducting inference conditional on a single selected model would ignore the model selection uncertainty, and often leads to serious consequences. A brief summary is provided as follows:
\begin{enumerate}
	\item There is a distortion between the estimator of the parameter of interest that takes account of the uncertainty in the model selection results and the one that does not (Zhang et al., \cite{Zhang2022psireview}). For instance, through a series of work (P\"otscher, \cite{potscher1991effects}; Leeb and P\"otscher, \cite{leeb2003finite,leeb2005model,leeb2006,leeb2008can}), they showed that the finite and asymptotic distributions of the post-model-selection parameter estimators, whether conditional on the selected model or not, generally deviate from Gaussian. These distributions are in fact mixtures of distributions, with each component distribution corresponding to a candidate model from the candidate model set. As such, the complicated forms of these distributions make it difficult to perform statistical inference, and often lead to overestimation of the precision (Allenbrand and Sherwood, \cite{Allenbrand2023}) and non-trivial biases of the resulting estimators (Chatfield, \cite{Chartfield1995}).
	\item Based on the distorted sampling distribution, the conditional (on the selected model) and unconditional coverage probabilities of the resulting naive confidence intervals are lower than their nominal levels, as shown both theoretically and experimentally by Hurvich and Tsai \cite{Hurvich1990}, Regal and Hook \cite{Regal1991}, Hoeting et al. \cite{Hoeting1999}, Hjort and Claeskens \cite{Hjort2003} and Zhang et al. \cite{Zhang2022psireview}. This result is also accompanied by the inflated Type I error rates (Berk et al., \cite{Berk2013}; Lee et al., \cite{Lee2016}; Tibshirani et al., \cite{Tibshirani2016exact}).
	\item The standard practice leads to researchers' overconfidence in the inferential results derived from the selected model, and underestimating the uncertainty associated with the parameter estimators (Hoeting et al., \cite{Hoeting1999}).
\end{enumerate}

\subsection{Methods to address model selection uncertainty}

In light of the above consequences, one should incorporate the impact of model selection uncertainty into the statistical inference results, as agreed by Buckland et al. \cite{buckland1997}. To this end, post-selection inference is a type of methods within the model selection framework that provides valid inference results by taking account of the effect of model selection uncertainty on statistical inference. Here, ``valid" means that the confidence interval for the parameter of interest $\theta_{\hat m}$, whether conditional on a selected model $\hat m$ or not, can achieve at least the nominal coverage probability $1-\alpha$ for $\alpha\in(0,1)$, regardless of the type of model selection methods. Zhang et al. \cite{Zhang2022psireview} and Kuchibhotla et al. \cite{Kuchibhotla2022} provide comprehensive reviews on the literature of post-selection inference, so in this section, we only give a brief review of this method in linear regression models and how it tackles the effect of model selection uncertainty on statistical inference.

To proceed the following, we first define basic settings. Consider a set of independently and identically distributed (i.i.d.) observations $\{y_i,x_i\}_{i=1}^n$, where $y_i$ and $x_i=(x_{i1},\ldots,x_{ip})^{\T}$ represent the $i$-th observation for the response variable and a $p$-dimensional covariates, respectively. Let ${Y}=(y_1, \ldots, y_n)^{\T} \in \mathbb{R}^n$ and $X=(x_1, \ldots, x_n)^{\T} \in \mathbb{R}^{n\times p}$. Ideally, the relationship between $Y$ and $X$ is modeled by a linear regression model with i.i.d. error terms $\epsilon=(\epsilon_1,\ldots,\epsilon_n)^{\T}$ and $\epsilon_i\sim N(0,\sigma^2)$, that is,
\begin{equation}\label{Con 12}
	Y=\mu+{\epsilon},
\end{equation}
where $\mu=X\beta^0$, and ${\beta^0}=(\beta_1^0, \ldots, \beta_p^0)^{\T}$ is termed as the true \textit{population-based regression coefficient} if equation \eqref{Con 12} is the data generating process (DGP) of $\{y_i,x_i\}_{i=1}^n$.

Let the index set ${M}=\{j_1, \ldots, j_m\} \subseteq\{1, \ldots, p\}$ correspond to a linear candidate model, which contains covariates with indices from $M$. Then, denote $X_M\in \mathbb{R}^{n\times m}$ as the design matrix of the candidate model. On this basis, this model $M$ can be built in a linear regression form $Y=X_M\beta_M+{\epsilon}$, based on which an ordinary least-squares (OLS) estimator for $\beta_M$ can be constructed as 
\begin{equation}\label{Con 13}
	\hat\beta_M=\left(X_M^{\T}X_M\right)^{-1}X_M^{\T}Y.
\end{equation}

Define $\mathcal{M}_{\text {all}}=\left\{{M}\mid {M} \subseteq\{1, \ldots, p\}, \text{rank}({X}_{M})=|M|\right\}$ as the model set containing all possible candidate models with full rank, and $\mathcal{M}_{\text {all}}$ may not necessarily contain the true model. Assuming that the design matrix ${X}$ is fixed, we define $\hat {M}_n(Y)$ to be a data-dependent model selection procedure, given by $\hat {M}_n(Y): \mathbb{R}^n \rightarrow\mathcal{M}_{\text {all}}$. For simplicity, we use $\hat {M}$ to represent the output of the map $\hat {M}_n(Y)$, i.e. $\hat {M}=\hat {M}_n({Y})$.

As the true DGP may not be a linear structure between $y_i$ and $x_i$, Berk et al. \cite{Berk2013} propose an alternative target to $\beta^0$, which is the \textit{projection-based target} $b_M$ that corresponds to a prespecified model $M$, and it is defined as
\begin{equation}\label{Con 14}
	b_M=\arg\min\limits_{b\in\mathbb{R}^{|M|}}\|\mu-X_Mb\|^2=\left(X_M^{\T}X_M\right)^{-1}X_M^{\T}\mu.
\end{equation}
The pros and cons of population-based regression coefficient and projection-based target have been discussed in detail by Berk et al. \cite{Berk2013} and Zhang et al. \cite{Zhang2022psireview}, the latter of which emphasizes that projection-based target is more applicable to the real-world setting, where it is difficult to verify the linearity of the underlying DGP. Therefore, focusing on the projection-based targets, we mainly introduce two post-selection inference methods that involves using $\hat\beta_M$ in \eqref{Con 13} to $b_M$. 

\subsubsection{Valid post-selection inference (PoSI)}

The PoSI method was proposed by Berk et al. \cite{Berk2013} and further extended by Bachoc et al. \cite{bachoc2020uniformly}, which is capable of producing universally valid post-selection confidence intervals for the projection-based elementwise targets $b_{j ,\hat {M}}, j \in \hat {M}$, regardless of model selection procedures and selected model $\hat {M}$. The merit of the PoSI method is that even if a selected submodel deviates from the true model, it still guarantees the nominal coverage probability $1-\alpha$ for that selected model, thus providing valid inference results. Now, we briefly introduce its framework.

Given a candidate model $\hat {M} \in \mathcal{M}_{\text {all}}$ selected by a generic model selection method $\hat {M}_n(Y)$, consider the following confidence interval for $b_{j , \hat {M}}$,
\begin{equation}\label{Con 15}
	\mathrm{CI}_{j , \hat {M}}(K)=\left(\hat{\beta}_{j , \hat {M}}-K \hat{\sigma}\left\{\left({X}_{\hat {M}}^{\T} {X}_{\hat {M}}\right)^{-1}\right\}_{j j}^{1 / 2}, \hat{\beta}_{j , \hat {M}}+K \hat{\sigma}\left\{\left({X}_{\hat {M}}^{\T} {X}_{\hat {M}}\right)^{-1}\right\}_{j j}^{1 / 2}\right),
\end{equation}
where $\hat{\beta}_{j , \hat {M}}$ is the $j$th element of $\hat\beta_{\hat M}$, $K$ is a constant to be introduced later, and $\{({X}_{\hat {M}}^{\T} {X}_{\hat {M}})^{-1}\}_{j j}$ is the $j$th diagonal element of $\{({X}_{\hat {M}}^{\T} {X}_{\hat {M}})^{-1}\}$. The variance $\sigma^2$ is estimated by $\hat{\sigma}^2=\text{SSE} /(n-p)$, where SSE is obtained under the full model. The intuition of this confidence interval generally comes from that of the $t$-test in regression analysis, where $K$ is replaced by $t(n-|\hat {M}| ; 1-\alpha / 2)$, i.e., the $(1-\alpha/2)$th percentile of a $t$-distribution with $n-\hat M$ degrees of freedom. 

On the basis of \eqref{Con 15}, PoSI aims to set a larger value of the constant $K$ in order to capture the uncertainty induced by the model selection procedure $\hat {M}_n(Y)$, and this goal is equivalent to finding the following confidence interval
\begin{equation}\label{Con 16}
	{P}\left(b_{j , \hat {M}} \in \mathrm{CI}_{j , \hat {M}}(K), \forall j \in \hat {M}\right) \geq 1-\alpha, 
\end{equation}
for any $\alpha \in(0,1)$.
In this way, the resulting confidence interval becomes wider and thus capable of achieving the nominal coverage probability, regardless of the choice of $\hat M$. To this end, Berk et al. \cite{Berk2013} propose the PoSI-constant $K_{\mathrm{PoSI}}$ as
\begin{equation*}
	K_{\mathrm{PoSI}}\left({X}, \mathcal{M}_{\text {all}}, \alpha, r\right)=
	\min \left\{K \in \mathbb{R}: P\left(\max _{M \in \mathcal{M}_{\text {all}}} \max _{j \in M}\left|t_{j, M}\right| \leq K\right) \geq 1-\alpha\right\}
\end{equation*}
where $r=n-p$, and
\begin{equation*}
	t_{j, M}=\frac{{e}_j^{\T}\left({X}_M^{\T} {X}_M\right)^{-1} {X}_M^{\T} {\epsilon}}{\hat{\sigma}\{({X}_M^{\T} {X}_M)^{-1}\}_{j j}^{1 / 2}},
\end{equation*}
with ${e}_j \in \mathbb{R}^{|M|}$ being the $j$th standard basis. Therefore, for any selected outcome $\hat {M}$, we have 
\begin{equation*}
	\max _{j \in \hat{M}}|t_{j ,\hat {M}}| \leq \max _{M \in \mathcal{M}_{\text {all}}}\max _{j \in M}\left|t_{j, M}\right|
\end{equation*}
which naturally leads to ${P}(\max _{j \in \hat {M}}|t_{j , \hat {M}}| \leq K_{\mathrm{PoSI}}) \geq 1-\alpha$. Hence, the target \eqref{Con 15} is satisfied by setting $K=K_{\mathrm{PoSI}}$. However, as $K_{\mathrm{PoSI}}$ is the $(1-\alpha / 2)$th percentile of $T=\max _{M \in \mathcal{M}_{\text {all}}} \max _{j \in M}\left|t_{j , M}\right|$, the latter of which is a random variable whose distribution depends on the error term $\epsilon$, we can approximate the distribution of $T$ by performing Monte Carlo experiments in order to calculate $K_{\mathrm{PoSI}}$.

Note that another choice for $K$ in \eqref{Con 15} is the Scheffe's constant (Scheffe, \cite{scheffe1999analysis}), that is given by $K_{\text {S}}=\sqrt{d \times F(d, r ; 1-\alpha)}$ under the linear regression model \eqref{Con 12}, where $d$ is the rank of the matrix ${X}$ and $F(d,r;1-\alpha)$ is the $(1-\alpha)$th percentile of an $F$-distribution with $d$ and $r$ degrees of freedom. As demonstrated in Berk et al. \cite{Berk2013}, $K_{\text {S}}$ is more conservative than $K_{\text {PoSI}}$ for all design matrices ${X}$ and any model set $\mathcal{M}_{\text {all}}$.

It is worth mentioning that the calculation of $K_{\text {PoSI }}$ is independent of the model selection process, as it does not involve any quantities of the selected model. However, due to the universal conservative nature of $K_{\text {PoSI }}$, the PoSI method is also necessarily conservative, since it provides a coverage guarantee for all potential selected outcomes. Another limitation of this method is its relatively high computational cost for computing the $K_{\text {PoSI }}$ due to the approximation process of the distribution of $T$ in face of a large number of candidate models. Therefore, this method is not applicable to the high-dimensional setting. Zhang et al. \cite{Zhang2022psireview} recommend the use of PoSI in the case of $p \approx 20$, and Kuchibhotla et al. \cite{Kuchibhotla2020} propose an improved version of the PoSI with computational efficiency. Finally, the PoSI requires the construction of the model set $\mathcal{M}_{\text {all}}$ before implementing the method, and this restriction prevents its use from sequential modeling, where the latter step relies on earlier steps, and $\mathcal{M}_{\text {all}}$ may evolve during this process.

\subsubsection{Exact post-selection inference (EPoSI)}
In contrast to the PoSI proposed by Berk et al. \cite{Berk2013}, Lee and Taylor \cite{lee2014exact}, Lee et al. \cite{Lee2016}, Hyun et al. \cite{Hyun2018} and Taylor and Tibshirani \cite{taylor2018post} propose another line of works that aim to construct confidence intervals with the exact $(1-\alpha)$ coverage probability. Another difference is that the EPoSI method depends on the type of model selection procedures and the selected model. Therefore, this method is known as the conditionally exact post-selection inference, and the resulting confidence interval for $b_{j,M}$, denoted by $\mathrm{CI}_{j,M}$, should satisfy
\begin{equation}\label{Con 17}
	P\left(b_{j , M} \in \mathrm{CI}_{j , M} \mid \hat {M}=M\right)=1-\alpha\; \text{for}\; j \in M.
\end{equation}

The EPoSI method generally contains five steps, which can be divided into two stages: (i) to determine the conditional distribution of $\hat{\beta}_{j , M} \mid \{\hat {M}=M\}$, which corresponds to Steps 1--3, and (ii) to construct the confidence intervals in \eqref{Con 17}, which involves Steps 4--5. Details of the EPoSI method is provided below.
\begin{enumerate}
	\item As the first stage intends to obtain the distribution of $\hat{\beta}_{j , M}$ conditional on $\{\hat {M}=M\}$, we first explore the distribution of $\hat{\beta}_{j , M}$ given the selection event
	\begin{equation}\label{Con 18}
		E_n(M, s)=\left\{{Y} \in \mathbb{R}^n: \hat {M}_n({Y})=M, \hat{s}_n({Y})=s\right\} =\{\hat {M}=M, \hat{s}=s\},
	\end{equation}
	where ${s}$ is the selected sign vector corresponding to $M$, $\hat{s}$ can be interpreted as the sign vector of the non-zero coefficient estimator under the model $\hat {M}$, and $\hat{s}_n(\cdot)$ is a map from $\mathbb{R}^n\to \{-1,1\}^{|M|}$.
	Note that $\hat{\beta}_{j , M}$ can be written as a linear form of $Y$, i.e., $\hat{\beta}_{j , M}={\eta}_{j,M}^{\T} {Y}$, where ${\eta}_{j,M}^{\T}={e}_j^{\T}\left({X}_{M}^{\T} {X}_{M}\right)^{-1} {X}_{M}^{\T}$. 
	Then, to investigate the exact conditional distribution of ${\eta}_{j,M}^{\T} {Y} \mid E_n(M, {s})$, it suffices to show that the selection event $E_n(M, {s})$ satisfies the \textit{polyhedral selection} property, that is,
	\begin{equation}\label{Con 19}
		E_n(M, s)=\left\{{Y} \in \mathbb{R}^n: {A}(M, s) {Y} \leq {B}(M, s)\right\},
	\end{equation}
	where ${A}(M, s)$ and ${B}(M, s)$ respectively denote an affine matrix and a vector that depend on the type of model selection procedure, the model $M$ and the sign vector $s$. The inequality in \eqref{Con 19} refers to the inequal elementwise relationship between two vectors. According to Lee et al. \cite{Lee2016}, several model selection procedures possess this polyhedral selection property, including lasso and elastic net, and the specific forms of ${A}(M, s)$ and ${B}(M, s)$ corresponding to these selection methods are available therein. 
	
	\item On the basis of \eqref{Con 19}, Lee et al. \cite{Lee2016} show that the polytope $\{{Y} \in \mathbb{R}^n: {A}(M, {s}) {Y} \leq{B}(M, s)\}$ can be further decomposed as
	\begin{eqnarray}\label{Con 20}
		\left\{{Y} \in \mathbb{R}^n: {A}(M, s) {Y}  \leq {B}(M, s)\right\} 
		&=&\big\{{Y} \in \mathbb{R}^n: V_s^{-}({Y})\leq {\eta}_{j,M}^{\T} {Y}\leq V_s^{+}(Y),\nonumber\\
		&&\;\;V_s^0(Y) \geq 0\big\}
	\end{eqnarray}
	where $V_s^-({Y}), V_s^+({Y})$ and $V_s^0({Y})$ are functions of $Y$ that are dependent on the type of model selection methods (see Lee et al. \cite{Lee2016} for their specific forms), and these quantities are shown to be independent of ${\eta}_{j,M}^{\T}{Y}$.
	
	\item Combining \eqref{Con 19} and \eqref{Con 20}, and the fact that ${Y} \sim {N}\left({\mu}, \sigma^2 {I}\right)$, we obtain
	\begin{equation}\label{Con 21}
		{\eta}_{j,M}^{\T} {Y} \mid\left\{{\eta}_{j,M}^{\T} {Y} \in\left[v_s^{-}, v_s^{+}\right]\right\} \sim \mathrm{TN}\left({\eta}_{j,M}^{\T} {\mu}, \sigma^2\left\|{\eta}_{j,M}\right\|^2 ; v_s^{-}, v_s^{+}\right)
	\end{equation}
	where $v_s^{-}, v_s^{+} \in \mathbb{R}$, and $\mathrm{TN}\left(\mu, \sigma^2 ; a, b\right)$ represents a Gaussian distribution with mean $\mu$ and variance $\sigma^2$ truncated to the interval $[a, b]$. So far, we have obtained the conditional distribution of $\hat{\beta}_{j , M}$ given $E_n(M,s)$, as in \eqref{Con 21}.
	
	\item Next, using the result in \eqref{Con 21}, we intend to obtain the distribution of $\hat{\beta}_{j , M}$ conditional on $\{\hat {M}=M\}$ and the confidence interval $\mathrm{CI}_{j , M}$ in \eqref{Con 17}. Let $F_{\mu, \sigma^2}^{[a, b]}(x)$ be the cumulative distribution function of $\mathrm{TN}\left(\mu, \sigma^2 ; a, b\right)$. By the Probability integral transformation, \eqref{Con 19} and \eqref{Con 21}, we have
	\begin{equation}\label{Con 22}
		\left\{F_{{\eta}_{j,M}^{\T} {\mu}, \sigma^2\left\|{\eta}_{j,M}\right\|^2}^{\left[V_s^{-}, V_s^{+}\right]}({\eta}_{j,M}^{\T} {Y}) \mid E_n(M, s)\right\} \sim \operatorname{Unif}(0,1)
	\end{equation}
	where $V_s^{-}=V_s^{-}({Y}), V_s^{+}=V_s^{+}({Y})$ and $\operatorname{Unif}(0,1)$ represents a continuous uniform distribution on the interval $(0,1)$.
	\item The pivot in \eqref{Con 22} enables us to construct the confidence intervals for $b_{j , M}$ in \eqref{Con 17}. Let $L_{j,s}$ and $U_{j,s}$ be two quantities satisfying $F_{L_{j,s}, \tau_j}^{\left[V_s^{-}, V_s^{+}\right]}({\eta}_{j,M}^{\T} {Y})=1-\alpha / 2$ and $F_{U_{j,s}, \tau_j}^{\left[{V}_s^{-}, {V}_s^{+}\right]}({\eta}_{j,M}^{\T} {Y})=\alpha / 2$. As $F_{\mu, \sigma^2}^{[a, b]}(x)$ is monotone decreasing in $\mu$ , we have
	\begin{equation}\label{Con 23}
		P\left(b_{j , M} \in\left[{L}_{j,s}, {U}_{j,s}\right] \mid E_n(M, s)\right)=1-\alpha.
	\end{equation}
	Next, considering the following decomposition for $\{\hat {M}=M\}$, 
	\begin{eqnarray*}
		\{\hat {M}=M\} & =&\bigcup_{s \in\{-1,1\}^{|M|}}\{\hat {M}=M, \hat{{s}}={s}\} \nonumber\\
		& =&\bigcup_{s \in\{-1,1\}^{|M|}}\left\{{Y}: {\eta}_{j,M}^{\T} {Y} \in\left[V_s^{-}({Y}), V_s^{+}({Y})\right], V_s^0({Y}) \geq 0\right\} \nonumber\\
		& =&\left\{{Y}: {\eta}_{j,M}^{\T} {Y} \in\left[\widetilde{V}^{-}({Y}), \widetilde{V}^{+}({Y})\right], \widetilde{V}^0({Y}) \geq 0\right\},
	\end{eqnarray*}
	we obtain the pivot uniform for all $M$, that is,
	\begin{equation}\label{Con 24}
		\left\{F_{\xi_j, \tau_j}^{\left[\tilde{V}^{-}, \tilde{V}^{+}\right]}({\eta}_{j,M}^{\T} {Y}) \mid \hat {M}=M\right\} \sim \operatorname{Unif}(0,1)
	\end{equation}
	by following the same argument as \eqref{Con 22}. On this basis, the lower and upper bound $L_j$ and $U_j$ of a confidence interval can be similarly constructed as that in \eqref{Con 23}. As such, the EPoSI confidence interval $\mathrm{CI}_{j , M}$ in \eqref{Con 17} is given by $\left[L_j, U_j\right]$.
\end{enumerate}

Compared to PoSI, the EPoSI method allows for the construction of confidence intervals to be dependent on the selection procedure as well as the selection outcome, while providing an exact $(1-\alpha)$ coverage guarantee for the resulting intervals. However, one potential drawback of the EPoSI is the lengths of its resulting confidence intervals derived from both \eqref{Con 22} and \eqref{Con 24}, because they are sometimes wider than certain competing confidence intervals, such as the OLS intervals, when the signal strengths are weak (Lee et al., \cite{Lee2016}), or may even have infinite lengths (Kivaranovic and Leeb, \cite{kivaranovic2021length}). Another limitation of this method is its computation cost. As it suffices to determine the interval $[V_s^-(Y),V_s^+(Y)]$ for each $s\in\{-1,1\}^{|M|}$, and the total number of these sign vectors is $2^{|M|}$, the computation cost can be enormous when $|M|$ is moderate or large. Finally, as pointed out by Kuchibhotla et al. \cite{Kuchibhotla2022}, it suffices to derive new theoretical analysis for the EPoSI method when switching to a new model selection procedure, while the PoSI is universally applicable to any model selection method.

\subsection{Model confidence set}\label{section 5}
In addition to the post-selection methods that aim to provide valid statistical inference results in the presence of model selection uncertainty, we introduce another type of methods, named model confidence set (MCS), that properly reflects the amount of useful information and thereby the uncertainty in the model selection procedure. 

By adapting the idea of confidence intervals for parameters, a class of methods, generally summarized as the model confidence set, aims to construct a set of candidate models that includes the best-performing model with a specified confidence level. This line of research is pioneered by Hansen et al. \cite{Hansen2011}, and later followed by Ferrari and Yang \cite{Ferrari2015}, Li et al. \cite{Li2019} and Zheng et al. \cite{Zheng2019}. Intuitively, the MCS procedure yields additional information about the uncertainty of model selection (Hansen et al., \cite{Hansen2011}).

The model confidence set is defined by Hansen et al. \cite{Hansen2011} as a set of candidate models that includes the best-performing model given a prespecified confidence level, and the criterion for measuring performance is subjectively determined by researchers. For better illustration, consider a model set $\mathcal{M}^0$ with cardinality $m_0$, and the $i$th model is evaluated by a loss function $L_{i,t}$ at time $t$. Then, the relative performance of the $i$th model compared to the $j$th model at time $t$ is given by
\begin{equation*}
	d_{ij,t}=L_{i,t}-L_{j,t},\quad i,j\in\mathcal{M}^0.
\end{equation*}
By assuming $\mu_{ij}=\E(d_{ij,t})$ to be independent of $t$, the $i$th model is considered to perform better than the $j$th model if $\mu_{ij}<0$. On this basis, define $\mathcal{M}^*=\{i\in \mathcal{M}^0:\mu_{ij}\leq 0 \text{ for all} j\in\mathcal{M}^0\}$, which represents the set of models with the best performances under the criterion $\{L_{i,t}\}_{i\in\mathcal M^0,t=1,\ldots,n}$. The target of the MCS is to construct a model set, denoted as $\widehat{\mathcal{M}}^*_{1-\alpha}$, that contains $\mathcal{M}^*$ with at least $(1-\alpha)$ probability either asymptotically or in a finite sample case. The construction of $\widehat{\mathcal{M}}^*_{1-\alpha}$ relies on an equivalence test $\delta_{\mathcal{M}}$ and an elimination rule $e_{\mathcal{M}}$, where $\delta_{\mathcal{M}}$ intends to test the hypothesis $H_{0,\mathcal{M}}:\mu_{ij}=0$ for all $i,j\in \mathcal{M}$,
with $\mathcal{M}\subset\mathcal{M}^0$ being the current model set to be examined. If $H_{0,\mathcal{M}}$ is rejected, then $\mathcal{M}$ is considered to include models that are not part of $\mathcal{M}^*$. In this case, the elimination rule $e_{\mathcal{M}}$ is employed to identify and remove the model with the worst performance from $\mathcal{M}$. The equivalence test $\delta_{\mathcal{M}}$ is then applied to the remaining model set, and this process is repeated until $H_{0,\mathcal{M}}$ is accepted. The remaining model set is then considered the MCS. The MCS is guaranteed to contain the best model with at least $(1-\alpha)$ probability. Moreover, as mentioned in Hansen et al. \cite{Hansen2011}, the size of the MCS reflects the amount of useful information contained in the data. When the data contains limited useful information, $\delta_{\mathcal{M}}$ may lack power, making it challenging to distinguish between good and poor models, which often leads to an early termination of the construction of the MCS and causes some of the poor models to be accidentally included in the MCS.

However, a drawback of the MCS is the interpretability of the remaining models therein, as it allows models with different structures. To this end, Ferrari and Yang \cite{Ferrari2015} utilize an $f$-test in a linear regression setting to develop the variable selection confidence set (VSCS) that is guaranteed to contain the true model with a prespecified confidence level. Similar to the MCS, the size of the VSCS can also be regarded as a reflection of the amount of useful information contained in the dataset. An insufficient amount of information results in a large number of models and higher uncertainty in the VSCS. Moreover, another use of the VSCS is to test whether a model is overly simplistic. If a model is excluded from the VSCS, it indicates the absence of certain important variables in this model, thereby being considered too parsimonious. Li et al. \cite{Li2019} propose the model confidence bounds (MCB), which aims to find a pair of nested models to provide confidence intervals for the true model. This MCB method provides more intuitive interpretations for its results, as the lower bound model can be regarded as the most parsimonious model that includes important covariates, while the covariates excluded by the upper bound model are considered redundant. Zheng et al. \cite{Zheng2019} form a model selection confidence set by likelihood ratio testing, which can contain all relevant information about the true model structure under certain conditions.

\section{Model selection instability}\label{section 4}
Some researchers believe that model uncertainty refers to model selection instability (Hong and Wang, \cite{Hongandwang2021}, Yuan and Yang, \cite{Yuan2005}, Chen et al., \cite{Chen2007}), which describes a situation where the model selection outcome is highly sensitive to the input data. In this case, even minor perturbations in the original data can lead to choosing a drastically different models, as some candidate models that differ from each other may end up sharing similar or even the same performances (Breiman \cite{Breiman1996,Breiman2001}, Chen et al., \cite{Chen2007}, Sun et al., \cite{Sun2023}, Chen et al., \cite{Chen2024timevarying}). This phenomenon is also noticed by Ding et al. \cite{Ding2018MSreview} and Zheng et al. \cite{Zheng2019}, the latter of which further points out the difficulty to demonstrate the superiority of a single model over its competitors due to the effect of the model selection instability. Such sensitivity poses a great challenge to the reliability of the resulting inferences. For instance, high instability can lead to unnecessarily large variance for the regression functions under the selected model (Yang, \cite{Yang2001}), and it is also responsible for impairing the reproducibility of the statistical findings (Ding et al., \cite{Ding2018MSreview}). If the initial model selection was highly unstable, findings based on that specific model might not be replicable in new samples drawn from the same population. Lastly, Yuan and Yang \cite{Yuan2005} suggest that a solid model selection process often exhibits small instability or uncertainty, thereby making more reliable selection results. On the contrary, if too much selection instability is involved, it is unlikely to identify the true model among the candidate models. Therefore, understanding and assessing model selection instability is crucial for evaluating the reliability of the subsequent statistical inference.

To measure the model selection instability, Yuan and Yang \cite{Yuan2005} develop the perturbation instability in estimation (PIE) criterion, which evaluates instability by introducing perturbations to the original data. For better illustration, consider the following setup
\begin{equation*} 
	Y_i=f(X_i)+\epsilon_i,\quad i=1,\ldots,n,
\end{equation*}
with the response $Y_i$, $d$-dimensional covariates $X_i$, the unknown regression function $f(\cdot)$, and the error term $\epsilon_i\sim N(0,{\sigma}^2)$. A number of $M$ linear candidate models $\left\{f_m\right\}_{m=1}^M$ are constructed by using different subset of $X_i$. Let $\hat{f}$ and $\hat{\sigma}$ be the estimates of $f$ and $\sigma$ conditional on the selected model and the original data $\left\{Y_i,X_i\right\}_{i=1}^n$. Define $\tilde{Y}_i=Y_i+W_i$ as the perturbed data, where $W_i\sim N(0,\rho^2\hat{\sigma}^2)$ is the perturbation term, with $\rho\in(0,1)$ reflecting the degree of perturbation. By repeating this perturbation process $B$ times, one obtains $B$ sets of perturbed datasets. For the $b$th set of data, the same model selection method is applied to identify the best performing model, denoted as $\tilde{f}_b$. On this basis, define
\begin{equation}\label{PIE}
	I(\rho)=\frac{1}{B} \sum_{b=1}^{B} \frac{\left[\sum_{i=1}^{n}\left\{\tilde{f}_{b}\left({X}_{i}\right)-\hat{f}\left({X}_{i}\right)\right\}^{2} / n\right]^{1 / 2}}{\hat{\sigma}}.
\end{equation}
It is evident that $I(\rho)$ quantifies the variation in the estimates of regression function based on the perturbed data and a model selection method. If $I(\rho)$ increases rapidly around $\rho=0$, then a minor perturbations can cause significant changes in the selected model, suggesting considerable instability in the model selection method. The slope of $I(\rho)$ at $\rho=0$ reflects the degree of instability in the specified model selection method. For this reason, PIE is defined as the slope of $I(\rho)$ at $\rho=0$, i.e., $I^{\prime}(\rho)|_{\rho=0}$. An interesting application of PIE is to decide when model averaging is more suitable than model selection. Specifically, Yuan and Yang \cite{Yuan2005} discover through experiments that the performances of model selection and model averaging often vary as the PIE increases. For instance, if the PIE values of the AIC and BIC exceed $0.5$, the predictive risk of the proposed model averaging method is lower than those selection results, and vice versa.

Chen et al. \cite{Chen2007} propose three different approaches for measuring model selection instability in the framework of factorial data analysis: parametric bootstrap instability, sequential instability and perturbation instability. These methods correspond to applying three perturbation methods to the original data, including imposing perturbation, sample deletion and resampling, and then measuring the level of instability based on the relative frequency of producing a different selected model based on the perturbed data to the selected model based on the original data. For instance, parametric bootstrap instability involves estimating the conditional mean of responses $\mu_{i_1\ldots i_\Phi}$ at each factor level $(i_1,\ldots,i_\Phi)$ and the variance of error terms $\sigma^2$ under a selected model. The corresponding estimators, $\hat{\mu}_{i_1\ldots i_\Phi}$ and $\hat{\sigma}^2$, are then utilized to generate a set of new observations for the response that follows a Gaussian distribution $N(\hat{\mu}_{i_1\ldots i_\Phi},\hat{\sigma}^2)$. This process is repeated many times. The same selection criterion is then applied to each new dataset to produce a selected model, and the relative frequency is then applied to measure the instability within the model selection process. It is evident that this relative frequency is related to the data generating process as well as the specific selection method that we adopt, and a high frequency value suggests great instability in the model selection method. For the rest two criteria, the perturbed approach used by parametric bootstrap instability is replaced with sample deletion and the resampling technique (Breiman \cite{Breiman1996}, Yuan and Yang \cite{Yuan2005}) for sequence instability and perturbation instability, respectively.
In addition, Chen et al. \cite{Chen2007} suggest that model selection instability can exert a distinct influence on the performances of model averaging/model selection methods. In cases where a model selection method exhibits high instability, model averaging typically outperforms model selection in terms of predictive risks, and vice versa. Therefore, model averaging is not inherently superior to model selection. This conclusion aligns with the findings of Yuan and Yang \cite{Yuan2005}.

\section{Summary}\label{section 6}
This paper provides a review of the existing literature on model uncertainty, and classifies the current understanding of model uncertainty into three different types: uncertainty about the true model, model selection uncertainty, and model selection instability. Notably, uncertainty about the true model is independent of any specific model selection method involved in the analysis, while the latter two interpretations are intrinsically related to specific model selection methods. As such, different selection procedures may yield different levels of uncertainty even for the same dataset according to these two interpretations. 

This paper further summarizes the methods proposed to address the three types of model uncertainty. For instance, both Frequentist model averaging and Bayesian model averaging are useful tools to perform statistical inference in face of the uncertainty about the true model. In addition, post-selection inference can provide valid confidence intervals for the parameter of interest, while taking account of the uncertainty and randomness of model selection. Finally, model averaging can serve as an alternative to model selection for prediction problems when the model selection process is relatively unstable.

Our final goal is to emphasize the importance of giving clear definitions of model uncertainty in researchers' papers, thereby preventing potential misunderstandings of this concept among readers. We expect that further research will lead to more studies on model uncertainty, contributing to a deeper understanding and exploration of this topic. This, in turn, should lead to the development of more effective methods for addressing the associated challenges.

\baselineskip=16pt
\bibliographystyle{unsrt}
\bibliography{ref}
\end{document}